\documentstyle[psfig,12pt]{article}
\begin{document}
\begin{center}
{\large\bf Random-field critical scattering at high magnetic\\
concentration in the Ising antiferromagnet $Fe_{0.93}Zn_{0.07}F_2$\\}
\vspace{0.25in}
{\large Z. Slani\v{c} $^a$, D.P. Belanger $^a$ and J.A. Fernandez-Baca $^b$\\}
\vspace{0.25in}
$^a$ {\small Department of Physics, University of California,
Santa Cruz, CA 95064 USA}\\
$^b$ {\small Solid State Division, Oak Ridge National Laboratory,
Oak Ridge, TN 37831-6393 USA}
\end{center}
\begin{abstract}
The high magnetic concentration Ising antiferromagnet $Fe_{0.93}Zn_{0.07}F_2$
does not exhibit the severe critical scattering hysteresis at low temperatures
observed in all lower concentration samples studied.  The system
therefore provides equilibrium neutron scattering line shapes suitable
for determining random-field Ising model critical behavior.
\end{abstract}
\vspace{0.4in}

It is well known that a phase transition
occurs \cite{by92} for the $d=3$ random-field Ising model (RFIM).
However, critical behavior investigations using the experimental
realizations, anisotropic randomly dilute
antiferromagnets such as $Fe_{x}Zn_{1-x}F_2$
in applied uniform fields, have been very challenging,
primarily as a result of severe hysteresis in
scattering line shapes below the transition $T_c(H)$.
This is most evident when comparing data taken
upon cooling in a field (FC) and
heating in the field after cooling in zero field (ZFC).
It has been shown \cite{bwshnlrl95,bkjn87}
that, at large dilution, the system breaks into
weakly interacting domains as $T \rightarrow T_c(H)$ 
after ZFC to establish long-range order.
For $x<0.8$, vacancies are so numerous
that domain walls form with little energy cost.
This effect obscures the decrease of the
order parameter within domains due to thermal fluctuations
since the Bragg intensity vanishes with domain formation.
In the present study, we have overcome this
problem by employing a magnetically concentrated crystal,
$Fe_{0.93}Zn_{0.07}F_2$, in which domains cannot form without
the energy penalty of breaking many magnetic bonds,
as in the original ferromagnetic RFIM.

The scattering measurements were made
at the Oak Ridge National Laboratory
High Flux Reactor using a two-axis spectrometer
configuration.  We used the (0 0 2) reflection
of pyrolytic graphite (PG) at an energy of $14.7$ meV
to monochromate the beam.  We employed three different
collimation configurations.  The lowest resolution
is with 70 min of arc before
the monochromator, 40 before the sample and
40 after the sample.  We also took scans
with 20 min of arc before and after the sample
and with 10 min of arc before and after the sample.
PG filters were used to eliminate
higher-order scattering.  The carbon thermometry scale was calibrated
to agree with recent specific heat results \cite{sb97} for the
$H=0$ transition.  The field dependence of the thermometry was also calibrated
and $T_c(H)-T_c(0)$ is consistent with the specific heat data.
The rounding of the transition occurs only for $|t|< 2 \times 10^{-3}$,
allowing the critical behavior to be probed reasonably well.

Four transverse ZFC scans across the antiferromagnetic
(1 0 0 ) Bragg point with fine collimation are shown in Fig. 1.
Two scans are at $0.13$ and $0.31$ K
above $T_c(H)=70.62K$ and the others are $0.14$ and $0.20$ K below.
The contrast between the scans above and below $T_c(H)$ is striking.
All scans taken at $H=0$ and those below $T_c(H)$ for $H>0$
can be reasonably fit to the mean-field Lorentzian line shape
\begin{equation}
S(q)= \frac{A} {q ^ 2 + \kappa ^ 2 } + {M_s}^2 \delta (q) \quad ,
\label{lor}
\end{equation}
where $\kappa (T)$ is the inverse correlation length for fluctuations
and $M_s$ is the staggered magnetization.  The mean-field RFIM prediction
includes a squared Lorentzian term
\begin{equation}
S(q)= \frac{A} {q ^ 2 + \kappa ^ 2 }
+  \frac{B} {(q ^ 2 + \kappa ^ 2 )^2 } + {M_s}^2 \delta (q) \quad ,
\label{lor_lor2}
\end{equation}
but the $H>0$ and $T<T_c(H)$ fits  allow no significant contribution from this
term.  Above $T_c(H)$ Eq.\ \ref{lor} does not fit the data well
and Eq.\ \ref{lor_lor2} is more appropriate.
The fitted behavior for the staggered susceptibility, $\chi (T)$,
and $\kappa (T)$, obtained from the Lorentzian term, does not behave
consistently as explained below.  We therefore emphasize that
these equations are not adequate, perhaps because
the RFIM exponent $\eta$, a measure of the deviation from
mean-field behavior, is quite large \cite{by92}.

The $q=0$ Bragg intensity is very large just below $T_c(H)$ and
vanishes just above, in accordance with Monte Carlo results \cite{r95}
that $\beta $ is close to zero.  The small
hysteresis observed in the Bragg intensity
upon FC and ZFC is consistent with the extinction effects.

The $|q|>0 $ line shapes also show a dramatic
change at $T_c(H)$ with no hysteresis.  For the previous studies
at $x=0.52$ \cite{bwshnlrl95} and $0.46$ \cite{bkjn87},
the $|q|>0$ intensities grow much more
rapidly as $T \rightarrow T_c(H)$ and, concomitantly,
the Bragg scattering exhibits an anomalous decrease in intensity,
effects attributed \cite{bwshnlrl95} to domain formation which are apparently
absent for $x=0.93$.

$\kappa (T)$ and $\chi (T)$ are shown in
Fig.\ 2 and Fig.\ 3, respectively, for both $H=0$ and $H=7T$.  In both cases
$\kappa (T)$ decreases to approximately the resolution
limit of $2 \times 10^{-3}$ reciprocal lattice units.
At $H=0$, the values of $\kappa (T)$ and $\chi (T)$ fit well to
random-exchange behavior \cite{bkj86} above and below $T_c(H)$,
with $\nu = 0.71 \pm 0.01$ and $\gamma = 1.35 \pm 0.01$, respectively.
For $H=7$~T, the values $\nu = 0.90 \pm 0.01$ and $\gamma = 1.72 \pm 0.02 $
for $T>T_c(H)$ are consistent with earlier
experiments \cite{bkj85}, simulations \cite{r95} and theory \cite{gaahs93}.
The disconnected staggered susceptibility exponent
$\bar{\gamma} = 3.0 \pm 0.1$ is also in reasonable agreement
with previous results \cite{bkj85}.  However, the values of $\kappa (T)$
and $\chi (T)$ for $T<T_c(H)$
are inconsistent with any power laws and fits are consequently
not shown.  The results for $\kappa (T)$ and $\chi (T)$
for $T<T_c(H)$ are very dependent on
the scattering line shapes used in the analysis
and further discussion will be deferred to a more extensive report.
The values for $\kappa (T)$ and $\chi (T)$ are independent of
the cooling procedure, implying equilibrium behavior.

This work has been supported
by DOE Grant No. DE-FG03-87ER45324
and by ORNL, which is managed by
Lockheed Martin Energy Research Corp. for the U.S. DOE
under contract number DE-AC05-96OR22464.

\newpage

\begin{figure}[t]
\centerline{\hbox{
\psfig{figure=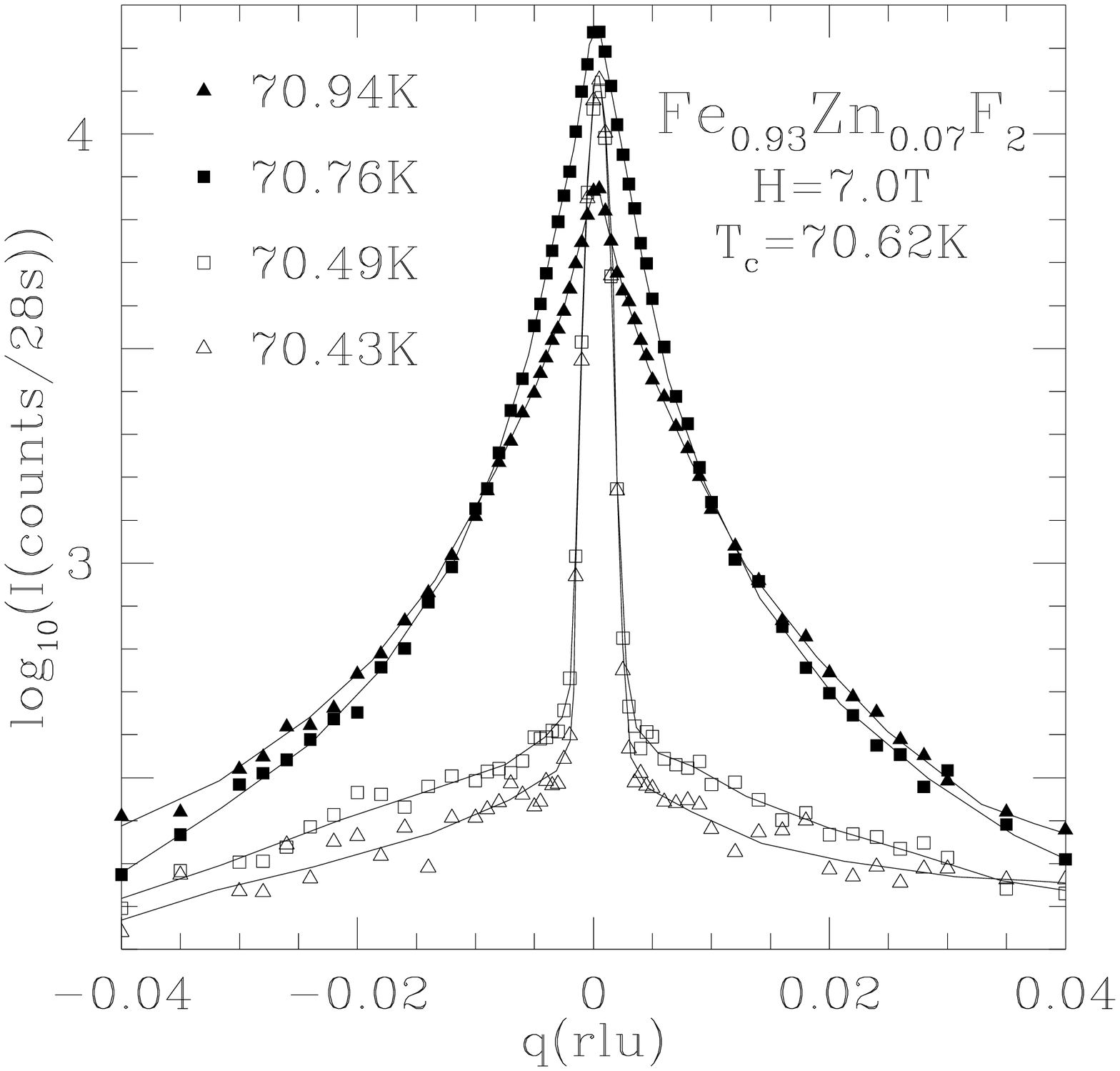,height=4.5in}
}}
\caption{The logarithm of the neutron scattering intensity
vs. $q$ just above and below $T_c(H)$.}
\end{figure}

\begin{figure}[t]
\centerline{\hbox{
\psfig{figure=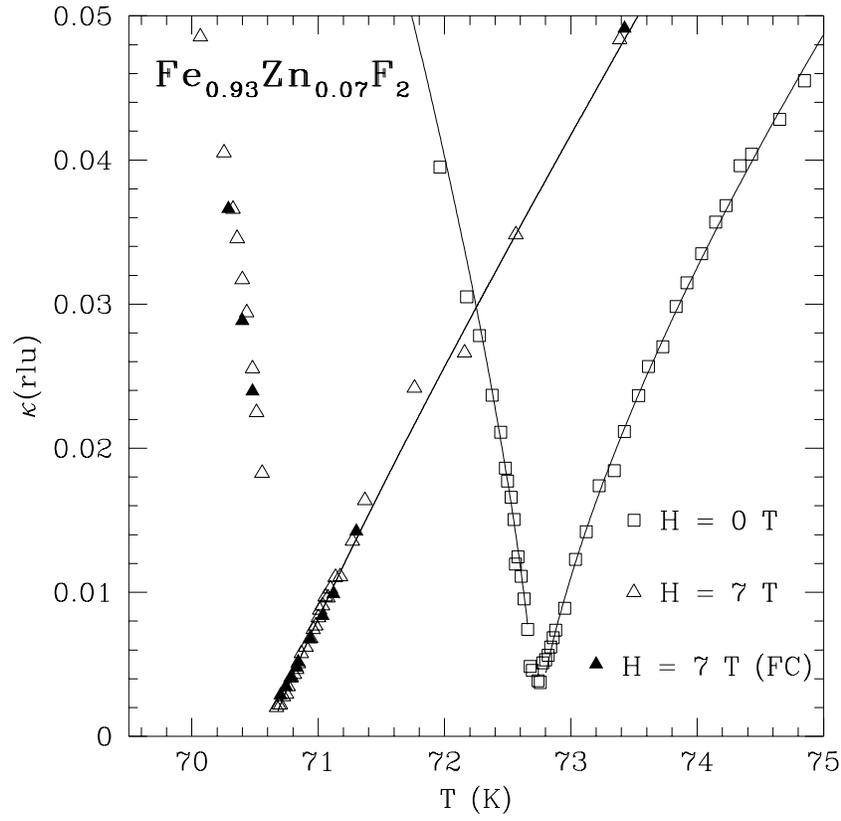,height=4.5in}
}}
\caption{The inverse correlation length $\kappa$ vs. T for
$H=0$ and $7$ T.}
\end{figure}

\begin{figure}[t]
\centerline{\hbox{
\psfig{figure=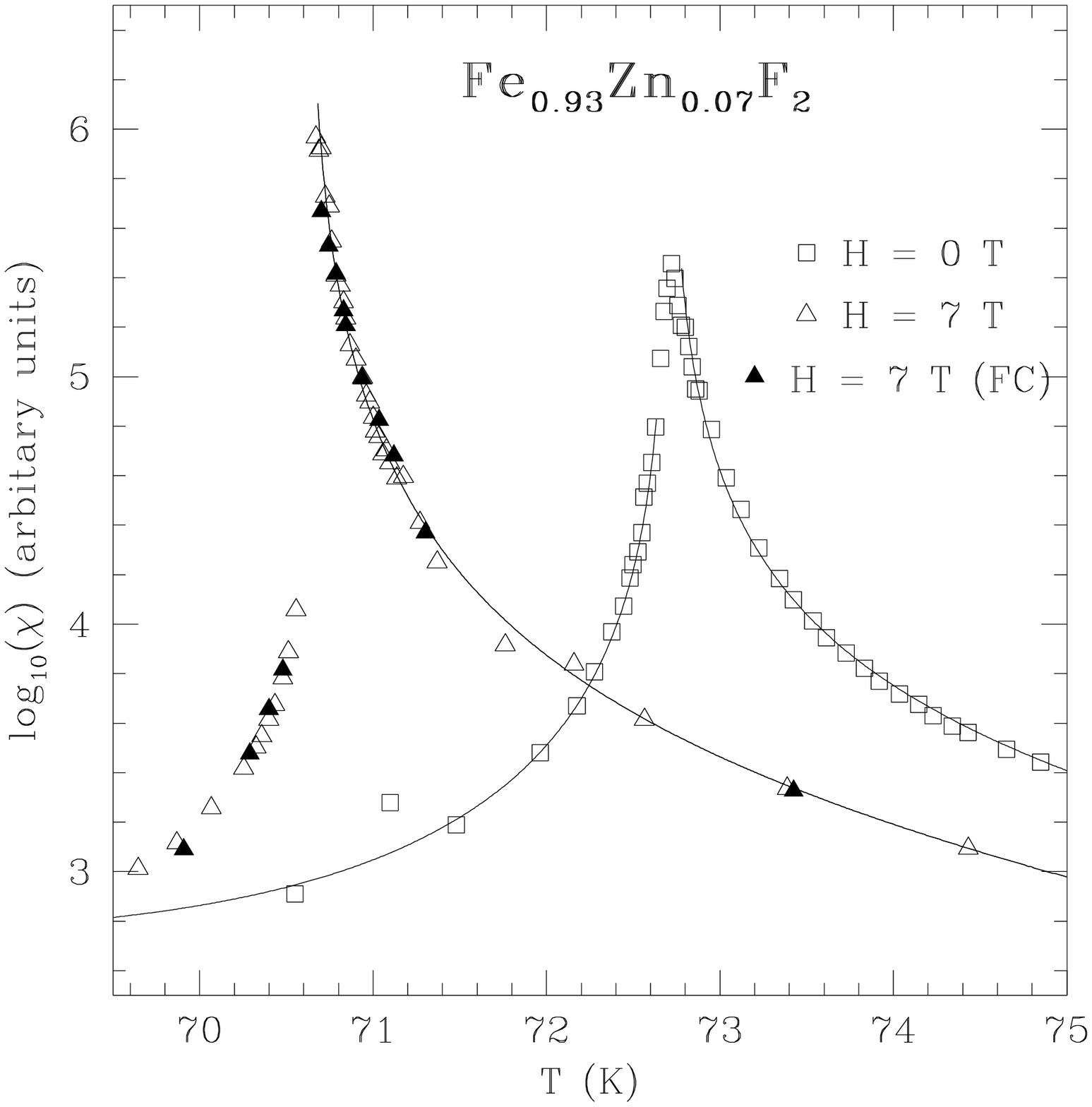,height=4.5in}
}}
\caption{The logarithm of the staggered susceptibility
$\chi$ vs. T for $H=0$ and $7$ T.}
\end{figure}


\begin{thebibliography}{999}

\bibitem{by92}For a review, see D.P. Belanger and A.P. Young,
J. Mag. Mag. Mater. 100 (1991) 272.

\bibitem{bwshnlrl95}D.P. Belanger, J. Wang, Z. Slani\v{c}, S.-J. Han,
R.M. Nicklow, M. Lui, C.A. Ramos, and D. Lederman,
J. Magn. Magn. Mater. 140-144 (1995) 1549;
Phys. Rev. B 54 (1995) 3420.

\bibitem{bkjn87}D.P. Belanger, A.R. King, V. Jaccarino, and R.M. Nicklow,
Phys. Rev. Lett. 59 (1987) 930.

\bibitem{sb97}Z. Slani\v{c} and D.P. Belanger,
submitted for publication, 1997.

\bibitem{r95}H. Rieger, Phys. Rev. B 52 (1995) 6659.

\bibitem{bkj86}D.P. Belanger, A.R. King and V. Jaccarino,
Phys. Rev. B 34 (1986) 452.

\bibitem{bkj85}D.P. Belanger, A.R. King and V. Jaccarino,
Phys. Rev. B 31 (1985) 4538.

\bibitem{gaahs93}M. Gofman, J. Adler, A. Aharony, A.B. Harris and M. Schwartz,
Phys. Rev. B 53 (1996) 6362.

\end{thebibliography}
\end{document}